\documentclass[twocolumn,amsmath,amssymb,aps,prl,floatfix,nofootinbib,superscriptaddress]{revtex4}
\usepackage{graphicx,graphics,times,color}
\usepackage{bm}
\usepackage{longtable}

\def\be{\begin{equation}}
\def\ee{\end{equation}}
\def\ba{\begin{eqnarray}}
\def\ea{\end{eqnarray}}
\def\la{\langle}
\def\ra{\rangle}

\begin{document}

\title{Entanglement Routers Using Macroscopic Singlets}

\author{Abolfazl Bayat}
\affiliation{Department of Physics and Astronomy, University College
London, Gower St., London WC1E 6BT, United Kingdom}

\author{Sougato Bose}
\affiliation{Department of Physics and Astronomy, University College
London, Gower St., London WC1E 6BT, United Kingdom}

\author{Pasquale Sodano}
\affiliation{Department of Physics, University
of Perugia, and INFN, Sezione di Perugia, Via A. Pascoli, 06123, Perugia, Italy}

\begin{abstract}
We propose a mechanism where high entanglement between very
distant boundary spins is generated by suddenly connecting two
long Kondo spin chains. We show that this procedure provides an
efficient way to route entanglement between multiple distant sites. 
We observe that the key features of the
entanglement dynamics of the composite spin chain are remarkably
well described using a simple model of two singlets, each formed
by two spins. The proposed entanglement routing mechanism is a
footprint of the emergence of a Kondo cloud in a Kondo system and can be engineered and observed
in varied physical settings.
\end{abstract}

\date{\today}
\maketitle

{\em Introduction:-} A high entanglement between two well
separated qubits is the central resource for quantum communication
tasks. Entanglement between arbitrary pairs
of distant qubits in a multi-qubit network enables the linking of
several quantum registers to a single larger computer. It also
facilitates the preparation
multi-particle entangled states \cite{Benjamin}, for measurement
based quantum computation. One could ask whether many-body systems
can serve as mediums for entanglement between arbitrary distant
qubits in a multi-site network. Though this is the most important
question from an ``applied" perspective, the thriving field of
entanglement in many-body systems
 \cite{Amico-RMP} remains focussed on the entanglement of blocks and proximal spins.
 In fact, long range entanglement between individual spins
is notoriously uncommon \cite{bose-vedral-fazio-osborne}. There
are proposals exploiting weak couplings of distant spins to a spin
chain \cite{lorenzo1-illuminati,lorenzo2-plenio-li-wojcik}, but
these have limited thermal stability or a very long time-scale of
entanglement generation. Alternatively, a global quench
\cite{Hannu-Sen} or specific time dependent couplings
\cite{Hangii-router} may generate entanglement, though this decays
with the system size. Finally, there is a proposal
\cite{bayat-quench} for distance independent entanglement through
a local quench which, however, lacks the versatility
of routing entanglement between multiple sites. A few quantum
routers have been recently proposed \cite{route}, but harnessing a
canonical many-body phenomenon for routing still is an open
question.

Kondo systems \cite{Sorensen-Affleck,kondo,bayat-kondo} are
very distinctive in the context of entanglement for at least two
reasons. Despite being ``gapless", they support the emergence
of a length scale $\xi$- the so called Kondo screening length
\cite{kondo,Sorensen-Affleck}- which can be tuned by varying only one parameter \cite{Sorensen-Affleck} and reflects in the
entanglement \cite{bayat-quench}, making it markedly different from other
conventional gapless models. Furthermore,
in Kondo systems, the impurity spin is maximally entangled \cite{bayat-kondo} with a block of spins whose spatial extent may be varied at will
by tuning $\xi$.

In this letter, we propose a dynamical mechanism by which long range
distance independent entanglement may be generated by the switch
on of a single coupling suddenly connecting two {\em macroscopic
singlets}. We show that this mechanism provides an
efficient way to route entanglement between various distant
parties.
By a macroscopic singlet we mean an arbitrarily long spin
chain which has been engineered to behave as a Kondo system of pertinent $\xi$ and thereby as a two spin singlet.
Indeed we show that, the key features of our mechanism for the two macroscopic singlets, are
remarkably well described by a four spin system made of two
singlets.

{\em Simple example:-} Let us first consider two spin singlets
each formed by only two spins interacting with a Heisenberg
interaction of strength $J'_1$ and $J'_2$, respectively. The
ground state of the composite system is then given by
$|gs\ra=|\psi^-\ra \otimes |\psi^-\ra$ with
$|\psi^-\ra=(|01\ra-|10\ra)/\sqrt{2}$. In this simple setting, one
may generate high entanglement between the boundary spins,
1 and 4 by merely turning on an interaction $J_m$ between the
spins 2 and 3. After quenching, the evolution of the system is
ruled by the Hamiltonian $H=J'_1
\vec{\sigma}_1.\vec{\sigma}_2+J'_2\vec{\sigma}_3.\vec{\sigma}_4+J_m\vec{\sigma}_2.\vec{\sigma}_3$
and, since the initial state is a global singlet, time evolution
allows for a nonzero overlap only with the singlet subspace of the
spectrum of $H$ so that
\begin{eqnarray}\label{psi_t1}
|\psi(t)\ra=e^{-iE_{S_1}t}|S_1\ra \la S_1|gs\ra+e^{-iE_{S_2}t}|S_2\ra \la S_2|gs\ra,
\end{eqnarray}
where, $|S_1\ra$ and $|S_2\ra$ are two singlet eigenvectors of $H$
with energy $E_{S_1}=-4J_m$ and $E_{S_2}=0$ respectively. In order to get
maximal entanglement between the boundary spins 1 and 4-- after a certain time $t^*$--
one has to choose $J_m=J'_1+J'_2$.
Once this condition is satisfied the state of the system at time $t$, up to a global phase, is given by
\begin{eqnarray}\label{psi_t2}
|\psi(t)\ra=\frac{-i\sin(2J_mt)}{2}(|0011\ra+|1100\ra)-\cr
 \frac{\cos(2J_mt)}{2} (|1001\ra+|0110\ra)+\frac{e^{i2J_mt}}{2}(|0101\ra+|1010\ra) .
\end{eqnarray}
Surprisingly, $|\psi(t)\ra$ depends only on $J_m$ and, by tracing
out the spins 2 and 3, one gets the density matrix $\rho_{14}(t)$
of the boundary spins. The entanglement between the spins 1
and 4 may be easily computed using concurrence \cite{wootters}
yielding
\begin{equation}\label{conc}
E=max\{0,\frac{1-3\cos(4J_mt)}{4}\}.
\end{equation}
Eq. (\ref{conc}) shows that $E$ oscillates with a period of
$\frac{\pi}{2J_m}$ and that, at time $t^*=\frac{\pi}{4J_m}$, the
spins 1 and 4 form a singlet state. In this simple setting one sees that: (i)
the entanglement dynamics is determined only by two singlet
eigenvectors of $H$; (ii) that maximal entanglement is achieved
only when $J_m=J'_1+J'_2$; (iii) the
entanglement dynamics is oscillatory with period $2t^*$, which is only a function of $J_m$ and,
thus, does not depend on $J'_1$ and $J'_2$ separately.

\begin{figure}
\centering
    \includegraphics[width=7cm,height=5cm,angle=0]{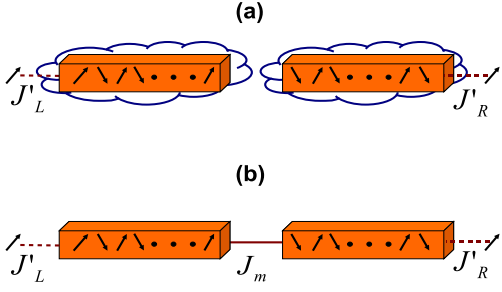}
    \caption{(Color online) (a) The composite system made of two separate Kondo spin chains initialized in their ground states in the Kondo regime.
    The extension of the clouds is tuned by $J'_R$ and $J'_L$ so that $\xi_k=N_k-1$ ($k=R,L$). (b) To induce dynamics, one switches on the interaction between the two chains by the amount $J_m$. }
     \label{fig1}
\end{figure}

{\em Generalization to a many-body system:-} We now show that the
above simple dynamics and the resulting high entanglement between
the boundary spins, may be reproduced even with many-body
systems-- for {\em arbitrary length scales}-- using pertinent spin
chains. For this purpose we consider two Kondo spin chains
\cite{Sorensen-Affleck} in the Kondo regime, i.e. two chains of
lengths $N_k$ described by
\begin{eqnarray}\label{HR_HN}
    H_{k}&=&J'_{k}(J_1\vec{\sigma}^k_1.\vec{\sigma}^k_2+J_2\vec{\sigma}^k_1.\vec{\sigma}^k_3)\cr
    &+&J_1\sum_{i=2}^{N_k}\vec{\sigma}^k_i.\vec{\sigma}^k_{i+1}+J_2\sum_{i=2}^{N_k-2}\vec{\sigma}^k_i.\vec{\sigma}^k_{i+2}, \ \ k=R, L
\end{eqnarray}
where, $J_1$ and $J_2$ are nearest and next to nearest neighbor couplings, $k=R$ ($k=L$) labels the right (left) chain, $\vec{\sigma}^k_i$ is the
vector of three Pauli operators at site $i$ for the chain $k$ and $J'_R$ ($J'_L$) is the impurity coupling of the right (left) hand side.

\begin{table}
\begin{centering}
\begin{tabular}{|c|c|c|c|c|c|c|c|c|c|}
  \hline
  $N_k$      & 4     & 6     & 8     & 10    & 12    & 14    & 16    & 18    & 20 \\
  \hline
  $J'_k$ & 0.300 & 0.280 & 0.260 & 0.250 & 0.240 & 0.230 & 0.220 & 0.215 & 0.210 \\
  \hline
\end{tabular}
\begin{tabular}{|c|c|c|c|c|c|c|c|c|c|}
  \hline
  $N_k$      & 22     & 24     & 26     & 28    & 30    & 32    & 34    & 36    & 38 \\
  \hline
  $J'_k$     & 0.205 & 0.202   & 0.198 & 0.195 & 0.190 & 0.187 & 0.184 & 0.180 & 0.175 \\
  \hline
\end{tabular}
\caption{Typical values of $J'_k$ to generate Kondo clouds of size
$\xi_k$ as given in Eq. (\ref{xi_RL}).}
\par\end{centering}
\centering{}\label{table_1}
\end{table}

It is well known that a Kondo spin chain supports a crossover from a gapless Kondo regime for $J_2<J_2^c=0.2412J_1$ to a gapped dimerized
regime for $J_2>J_2^c$.
In the Kondo regime the Kondo screening length is uniquely determined by
the impurity coupling \cite{Sorensen-Affleck,bayat-kondo} and, for large chains, the explicit dependence is given by $\xi_k=e^{\alpha/\sqrt{J'_k}}$,
where $\alpha$ is a constant; $\xi_k$ sets the size of a block of spins forming a singlet with the impurity \cite{bayat-kondo}. In the following we
shall fix the value of $J'_R$ and $J'_L$ so that
\begin{equation}\label{xi_RL}
\xi_k=N_k-1, \ \ \ k=R, L.
\end{equation}
We report in table I the values of the impurity couplings--
determined for chains of arbitrary lengths in Ref.
\cite{bayat-kondo}-- as $N_k$ is increased. Eq. (\ref{xi_RL})
allows to build two macroscopic singlets (i.e., extended over a
distance $\xi_k$ tuned by $J'_k$). The composite spin system is
depicted in Fig. \ref{fig1}(a); the two impurities sit at the
opposite sides and may be regarded as the boundary spins of the
composite system while, due to Eq. (\ref{xi_RL}), the two Kondo
clouds are tuned to take over each chain separately. Note that not only is this $J'_k \sim 1/Log^2N_k$ much stronger than the weak couplings in Refs. \cite{lorenzo1-illuminati,lorenzo2-plenio-li-wojcik}, but also the chain is {\em gapless}, so it cannot lead to perturbative end-to-end effective Hamiltonians.

Initially, the two chains are separated and initialized in their
ground states (see Fig. \ref{fig1}(a)) and the initial state of
the composite chain is given by $|\psi(0)\ra=\Pi_{k=R,L}|GS_k\ra$
where $|GS_k\ra$ is the ground state of the chain $k$. Then, we
switch on
\begin{equation}\label{HR_HN}
    H_I=J_m (J_1\vec{\sigma}^L_{N_{_L}}.\vec{\sigma}^R_{N_{_R}}+J_2\vec{\sigma}^L_{N_{_L}-1}.\vec{\sigma}^R_{N_{_R}}+
    J_2\vec{\sigma}^L_{N_{_L}}.\vec{\sigma}^R_{N_{_R}-1}).
\end{equation}
between the two chains (see Fig. \ref{fig1}(b)). The Hamiltonian
of the composite system of length $N=N_L+N_R$ is given by
$H=H_L+H_R+H_I$. Now the ground state evolves according to
$|\psi(t)\ra=e^{-iHt}|\psi(0)\ra$. From  knowing $|\psi(t)\ra$ one
obtains the reduced density matrix of the boundary spins at a
generic time $t$ by tracing out all other spins from the state
$|\psi(t)\ra$ and evaluate the concurrence $E(t,J_m)$ between the
boundary spins. The dynamics is now not analytically
solvable and one has to resort to numerical simulations which, for
$N>20$, use the time dependent density matrix renormalization
group (tDMRG) introduced in \cite{whiteDMRG} while, for $N<20$,
one may use exact diagonalization. For temperatures $T<1/\xi\sim2/N$,
i.e. when the two constituent chains are in the Kondo ground state, we
find that the evolution of the composite chain well
reproduces all the relevant features exhibited by the simple
dynamics of a four spin system made out of two singlets.

If the composite system built out of two extended Kondo singlets should reproduce the remarkable features of the simple example discussed above one
should expect that, $E(t,J_m)$ oscillates with a period depending only on $J_m$ and that maximal entanglement between the boundary spins is reached
at the half of the period provided that
\begin{equation}\label{Jm_singlet}
J_m=\Phi(N)(J'_L+J'_R).
\end{equation}
$\Phi(N)$ accounts for the effects arising due to the
extended size of the Kondo singlets. Of course, for our dynamics
to make sense at all one has to require that, as $N \rightarrow
\infty$, $J_m$ should take a nonzero and finite limiting value $J$
(otherwise, one injects either zero or infinite energy). It is
remarkable that this condition alone suffices to determine
$\Phi(N)$. Indeed, if $N_L\approx N_R \rightarrow \infty$, one
has
\begin{equation}\label{phi_N}
\Phi(N)\sim \frac{J}{\alpha^2} \log^2(\frac{N}{2}),
\end{equation}
since, in the Kondo regime, one
has that $\xi_k=e^{\alpha/\sqrt{J'_k}}$.

\begin{figure}
\centering
    \includegraphics[width=8cm,height=5.6cm,angle=0]{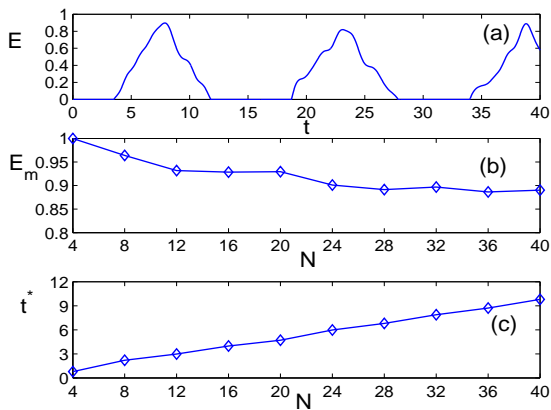}
    \caption{(Color online) (a) The oscillatory dynamics of entanglement $E$ vs. time $t$ in the Kondo regime ($J_2=0$) for a composite
    system of $N=32$ when $N_L=N_R$. (b) The maximal entanglement $E_m$ vs. $N$ at $t=t^*$ when $J_m$ has its optimal value.
    (c) $t^*$ vs. $N$.}
     \label{fig2}
\end{figure}

For the time being we will consider only chains for which
$N_L=N_R$ and so that $N=2N_L$. In Fig. \ref{fig2}(a) we plot the
evolution of the entanglement as a function of time for
$J_m=0.97J_1$ when $N=32$ in the Kondo regime ($J_2=0$) of each
chain. We see that entanglement dynamics is oscillatory with a
period $2t^*$. Restricting only to the first period of
oscillations one sees that, there is an optimal value of $J_m$ for
which, at time $t^*$, the entanglement reaches its maximum $E_m$.
In Fig. \ref{fig2}(b) we plot $E_m$ as a function of $N$.
Though the entanglement decreases as $N$ increases for short
chains, its value remains very high and becomes {\em distance
independent} for very long chains.  It is remarkable that this distance independent value seems to be
0.9 (e.g. for chains of length $N=40$) whereas for the only other distance independent case so far \cite{bayat-quench} it was merely 0.7. To
complete the picture of entanglement evolution in Fig.
\ref{fig2}(c) we plot $t^*$ as a function of $N$. We see that the
time needed to generate the entanglement between the boundary
spins increases linearly with $N$ with a slope that is small
enough to allow for fast dynamics. The linear dependence of $t^*$
on $N$ implies that, for a system composed of two extended Kondo
singlets, $t^*$ is related to $J_m$ by
\begin{equation}\label{topt_N} t^*\sim N \sim \xi_k \sim
e^{\alpha\sqrt{\frac{2\Phi(N)}{J_m}}}.
\end{equation}

\begin{figure}
\centering
    \includegraphics[width=8cm,height=5.6cm,angle=0]{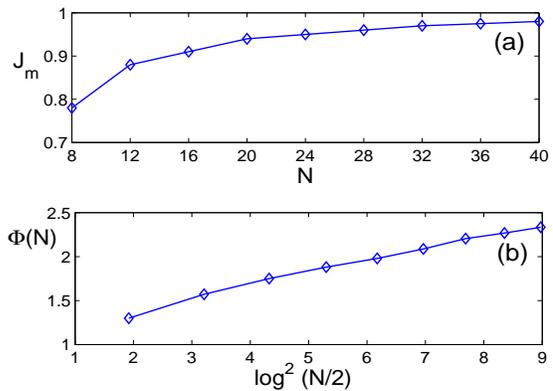}
    \caption{(Color online) (a) Optimal $J_m$ vs. $N$ for $N_L=N_R$ in the Kondo regime. (b) The asymptotic behavior
    of $\Phi(N)$ vs. $\log^2(\frac{N}{2})$.}
     \label{fig3}
\end{figure}

In Fig. \ref{fig3}(a) we have plotted the optimal value of $J_m$
as a function of $N$. One sees that, as $N$ increases, $J_m$ goes
to 1 thus, confirming the assumption used in the derivation of
$\Phi(N)$ (see Eq. (\ref{phi_N})). In Fig. \ref{fig3}(b) we plot
$\Phi(N)$ versus $\log^2(N/2)$. The linearity of the plot provides
an independent numerical confirmation of the result
obtained in Eq. (\ref{phi_N}).

\begin{figure}
\centering
    \includegraphics[width=8cm,height=5.6cm,angle=0]{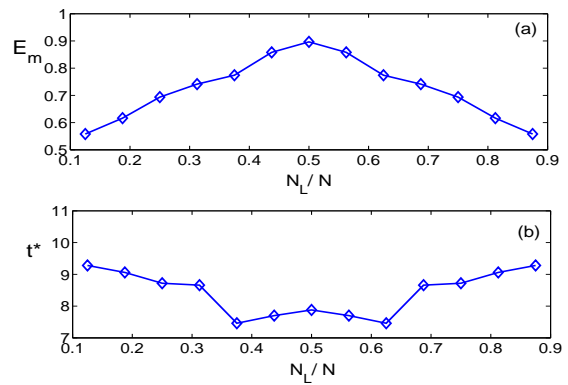}
    \caption{(Color online) Non-symmetric case where $N=32$ and $N_L$ and $N_R$ are varied: (a) $E_m$ vs. $N_L/N$.
    (b) $t^*$ vs. $N_L/N$.  }
     \label{fig4}
\end{figure}

Our numerical approach allows to investigate also situations for
which $N_L$ is different from $N_R$. In Fig. \ref{fig4}(a) we plot
$E_m$ versus $N_L/N$ for a composite chain of length $N=32$. Fig.
\ref{fig4}(a) shows that the entanglement is maximal when
$N_L=N_R$ and that decreases sensibly when the sizes of the
constituent Kondo chains are very different. In Fig. \ref{fig4}(b)
we plot $t^*$ as a function of $N_L/N$. Again one sees that the
optimal time $t^*$ is much shorter when
$N_L\sim N_R$. Figs. \ref{fig4}(a) and (b) lead us to conclude
that efficient routing of entanglement is possible only if
$N_L\sim N_R$.

\begin{table}
\begin{centering}
\begin{tabular}{|c|c|c|c|c|c|c|c|c|c|}
  \hline
  $N$      & 8     & 12    & 16    & 20    & 24    & 28    & 32    & 36    & 40 \\
  \hline
  $E_m$(K) & 0.964 & 0.932 & 0.928 & 0.929 & 0.901 & 0.891 & 0.897 & 0.886 & 0.891 \\
  \hline
  $E_m$(D) & 0.957 & 0.903 & 0.841 & 0.783 & 0.696 & 0.581 & 0.468 & 0.330 & 0.160 \\
  \hline
  $t^*$(K) & 2.200 & 2.980 & 3.980 & 4.700 & 5.980 & 6.800 & 7.880 & 8.720 & 9.800 \\
  \hline
  $t^*$(D) & 3.780 & 7.290 & 10.32& 13.41& 16.89& 20.43& 24.51& 27.12& 35.01 \\
  \hline
\end{tabular}
\caption{Comparison between $E_m$ and $t^*$ for a Kondo spin chain
in the Kondo ($J_2=0$) and dimer regimes ($J_2=0.42$). In the table $K$ stands for Kondo
and $D$ for dimer. }
\par\end{centering}
\centering{}\label{table_1}
\end{table}

The proposed mechanism for generating high entanglement between
the boundary spins of a composite spin system relies heavily on
Eq. (\ref{xi_RL}) and, thus, on the fact that, for Kondo chains of
arbitrary sizes $N_L$ and $N_R$, one can always tune the impurity
couplings $J'_L$ and $J'_R$ so as to make the Kondo cloud
comparable with the size of the chains. As a result, entanglement
generation between the boundary spins should vanish for
$\xi_k<N_k/2$ as well as when the constituent Kondo chains
are in the dimer regime (i.e., $J_2>J_2^c$) where the cloud does
not exist at all. We computed numerically $E_m$ and $t^*$,
for a chain composed of two Kondo spin chains in the dimer
regime. The results are reported in table II and compared with
the results obtained for the same quantities when the two constituent chains are in the Kondo regime and Eq. (\ref{xi_RL})
is satisfied. Table II shows that, as $N$ increases, entanglement $E_m$
(optimal time $t^*$) is very small (large): for instance, for $N=40$, in the dimer regime, $E_m=0.16$ and $t^*=35.01$ while, in
the Kondo regime, $E_m=0.89$ and $t^*=9.80$.

\begin{figure}
\centering
    \includegraphics[width=7cm,height=5cm,angle=0]{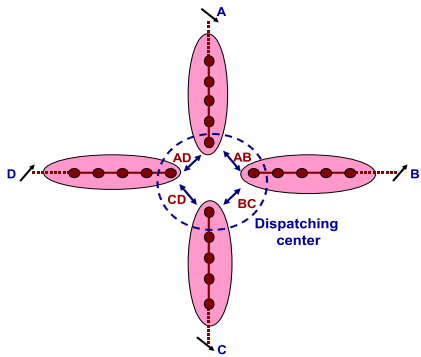}
    \caption{(Color online) A four-node  router in which each user controls one boundary spin. A dispatcher connects two chains to induce
    dynamics in a channel composed of two spin chains in order to generate entanglement between the boundary spins.}
     \label{fig5}
\end{figure}

{\em Entanglement router:-} Our analysis allows to engineer an
efficient entanglement router dispatching entanglement between
very distant qubits. A four-node entanglement router is sketched
in Fig. \ref{fig5}. Each node, say $A$, $B$, $C$ and $D$, has a
boundary spin whose coupling to its adjacent chain is tuned so as
to generate a Kondo cloud reaching the dispatch center (Fig.
\ref{fig5}). The dispatcher can entangle the spins of two
arbitrarily chosen nodes, say $A$ and $B$, by switching on a
coupling $J_m$ between the chains $A$ and $B$ and, thus, induce
the quench dynamics previously analyzed. At $t=t^*$, the
entanglement may be taken out of the boundary spins by a fast swap
to any memory qubits in nodes $A$ and $B$ for building
resources for quantum computation. Note that exclusive
pairs of nodes, e.g. $(A,B)$ and $(C,D)$, can be connected
simultaneously.


{\em Implementations and challenges:-} Spin chains with switchable/tunable couplings are realizable \cite{impl} with both superconducting qubits and spins in quantum dots.
In the former, the effect of a reasonable dephasing of strength $0.005J_1$ \cite{Ripoll} for $N=12$ is about $10\%$.
In the latter, a magnetic field in a random direction acts on each spin due to the dot nuclei \cite{taylor07}. Our simulations show that for $N=12$, a very strong magnetic field ($\sim 0.05J_1$) \cite{taylor07} decreases the entanglement by $5\%$.

 {\em Conclusions:-} We proposed a mechanism for generating high entanglement between distant spins by switching
 on an appropriate interaction between two Kondo spin chains. In contrast to other recent networking schemes \cite{route} it does not demand control of the intermediate spins or time-varying local fields. Our results hint that a Kondo spin chain
 satisfying Eq. (\ref{xi_RL}) may be effectively described by an extended singlet formed by two spins since the key features of the entanglement
 dynamics can be easily understood using a simple model of a pair of two spin singlets. Indeed, in this non-solvable model, the above is the best explanation of the strikingly high entanglement. From table II one sees that in the absence of the Kondo cloud, entanglement is suppressed; thus, the remarkable dynamical behavior of the system is a new clear footprint of the emergence of the Kondo cloud in a Kondo system.

 {\em Acknowledgements:-} Discussions with I. Affleck, H. Johanneson, N. Laflorencie and E. Sorensen  are  acknowledged.
 AB and SB (P.S.) thank(s) the University of Perugia (UCL) for hospitality and partial support.
 SB and AB acknowledge the EPSRC, and the Royal Society and the Wolfson Foundation.

\end{document}